\documentclass[11pt]{article}
\usepackage{graphicx}
\usepackage{subfig}
\usepackage{flafter}

\title{Rise of the Rest: The Growing Impact of Non-Elite Journals}

\author{
Anurag Acharya, Alex Verstak, Helder Suzuki, Sean Henderson,\\
Mikhail Iakhiaev, Cliff Chiung Yu Lin,  Namit Shetty\\
\\
Google Inc.
}

\begin{document}

\maketitle

\begin{abstract}

In this paper, we examine the evolution of the impact of non-elite
journals. We attempt to answer two questions. First, what fraction of
the top-cited articles are published in non-elite journals and how has
this changed over time. Second, what fraction of the total citations
are to non-elite journals and how has this changed over time. 

To answer these questions, we studied citations to articles published
in 1995-2013. We computed the 10 most-cited journals and the 1000
most-cited articles each year for all the 261 subject categories
included in Scholar Metrics. We considered the 10 most-cited journals
in a category as the {\it elite} journals for the category and all
other journals in the category as {\it non-elite}.

There are two main conclusions from our study. First, the fraction of
highly-cited articles published in non-elite journals increased
steadily over 1995-2013. While the elite journals still publish a
substantial fraction of high-impact articles, many more authors of
well-regarded papers in a diverse array of research fields are
choosing other venues.

Our analysis indicates that the number of top-1000 papers published in
non-elite journals for the representative subject category went from
149 in 1995 to 245 in 2013, a growth of 64\%. Looking at broad
research areas, 4 out of 9 broad areas saw at least one-third of the
top-cited articles published in non-elite journals in 2013. All broad
areas of research saw a growth in the fraction of top-cited articles
published in non-elite journals over 1995-2013. For 6 out of 9 broad
areas, the fraction of top-cited papers published in non-elite
journals for the representative subject category grew by 45\% or more.

Second, now that finding and reading relevant articles in non-elite
journals is about as easy as finding and reading articles in elite
journals, researchers are increasingly building on and citing work
published everywhere. Considering citations to all articles, the
percentage of citations to articles in non-elite journals went from
27\% of all citations in 1995 to 47\% in 2013. Six out of nine broad
areas had at least 50\% of total citations going to articles published
in non-elite journals in 2013.

\end{abstract}

\section{Introduction}

Several factors have driven the sustained impact of elite journals.
First, these journals have traditionally been available in many more
academic libraries worldwide. The costs of physical distribution and
storage required due to print publication meant only selected journals
would be widely available. Articles published in these journals had a
much higher likelihood of being read, built upon and cited.

Second, success metrics in scholarly communication had largely been
computed at the journal-level. Those of us who have been on the
academic job market well know that our resumes were likely to be
summarized as ``X articles in the top journals''. As a result,
researchers usually target elite journals for their best work.

Third, literature research approaches had primarily been either
browsing journal issues or scanning reverse chronological search
results (most-recent-first). These approaches present researchers with
a large number of articles to scan and require substantial effort to
track down relevant articles. As a result, researchers had been more
likely to limit the scope of their literature search to elite
journals.

There have been several dramatic changes in scholarly communication
over the last two decades that have the potential to significantly
influence these factors. First, scholarly journals have largely moved
from physical distribution of print issues to online availability of
individual articles. A large number of journals have also digitized
older articles and made them available online. Many publishers and
aggregators provide large collections as a part of Big Deal
licenses. As a result, it is easier for many more libraries to provide
access to publications beyond a core collection of elite journals.

Second, success metrics for researchers have expanded to include
article-level metrics. These include per-article citation counts as
well as aggregate metrics such as the {\it
  h-index}~\cite{hirsch2005index}. Furthermore, these metrics are
widely available to all users without a subscription -- which makes it
easier for both researchers and those considering their resumes to
view these metrics. This allows researchers to highlight the success
of their impactful articles no matter where they are published.

Third, search services now cover all available journals, instead of a
selected subset. Furthermore, they index the entire text of articles
instead of just abstracts and keywords. The common ranking approach
has moved from reverse chronological to relevance ranking
(most-relevant-first). Finding and reading relevant articles in
non-elite journals is now about as easy as finding and reading
articles in elite journals.

To understand the influence of these changes on the impact of
non-elite journals, we studied citations to articles published in
1995-2013. We attempted to answer two questions. First, what fraction
of the top-cited articles are published in non-elite journals and how
has this changed over time. This covers the impact of the most visible
papers, which are often the ones that make key contributions.  Second,
what fraction of the total citations to all articles are to non-elite
journals and how has this changed over time.  This covers the impact
of all papers.

We computed the 10 most-cited journals and the 1000 most-cited
articles each year for all the 261 subject categories included in
Scholar Metrics~\cite{suzuki2014}. We considered the 10 most-cited
journals in a category as the {\it elite} journals for the category
and all other journals in the category as {\it non-elite}.

There are two main conclusions from our study. First, the fraction of
highly-cited articles published in non-elite journals increased
steadily over 1995-2013. While the elite journals still publish a
substantial fraction of high-impact articles, many more authors of
well-regarded papers in a diverse array of research fields are
choosing other venues.

Our analysis indicates that the number of top-1000 papers published in
non-elite journals for the representative subject category went from
149 in 1995 to 245 in 2013, a growth of 64\%. Looking at broad
research areas, 4 out of 9 broad areas saw at least one-third of the
top-cited articles published in non-elite journals in 2013. All broad
areas of research saw a growth in the fraction of top-cited articles
published in non-elite journals over 1995-2013. For 6 out of 9 broad
areas, the fraction of top-cited papers published in non-elite
journals for the representative subject category grew by 45\% or more.

Second, now that finding and reading relevant articles in non-elite
journals is about as easy as finding and reading articles in elite
journals, researchers are increasingly building on and citing work
published everywhere. Considering citations to all articles, the
percentage of citations to articles in non-elite journals went from
27\% of all citations in 1995 to 47\% in 2013. Six out of nine broad
areas had at least 50\% of total citations going to articles published
in non-elite journals in 2013.

\section{Methods}

For this study, we included all journals and conferences that were
assigned to one or more subject category in the 2014 release of
Scholar Metrics. The Scholar Metrics inclusion criteria for
publication venues were~\cite{scholar-metrics2014}: (1) publish 100 or
more articles over 2009-2013, (2) at least one article must receive at
least one citation over 2009-2013, (3) follow Google Scholar indexing
guidelines. Scholar Metrics limits categorization into subject categories
to English publications.  Accordingly, this study covers all the
English language journals and conferences included in Scholar
Metrics. Scholar Metrics also includes selected preprint
repositories. Preprint repositories are not included in this study.

We used the subject categories from the 2014 release of Scholar
Metrics. We created a group of articles for each subject-category-year
combination, such as {\it Immunology} for the year 2000. Each
category-year group included all articles published in the given year
in all publications in the given category. 

For each publication, we included all articles with a publication date
within 1995-2013, both inclusive. Note that each journal or conference
can be associated with more than one subject category. Such
publications are included in the computation for each category they
are a part of.

We identified the 10 most-cited journals for each category-year group.
For this, we used the ordering mechanism used in Scholar
Metrics~\cite{scholar-metrics2014}: journals are sorted by their {\it
  h5-index} with ties being broken by their {\it
  h5-median}.\footnote{The {\it h5-index} of a publication is the
  largest number {\it h} such that at least {\it h} articles published
  in the last five complete calendar years in that publication were
  cited at least {\it h} times each. The {\it h5-core} of a
  publication is a set of top cited {\it h} articles from the
  publication that were published in the last five complete calendar
  years. These are the articles that the {\it h5-index} is based
  on. The {\it h5-median} of a publication is the median of the
  citation counts in its {\it h5-core}. The {\it h5-median} is a
  measure of the distribution of citations to the articles in the {\it
    h5-core}.  } These journals were considered as the {\it elite}
journals for the group. All other journals in the group were
considered {\it non-elite}. Note that the list of elite journals was
recomputed each year which allowed the analysis to capture changes in
the focus of subject categories as well as newly successful journals.

Next, we computed the list of the 1000 most-cited articles in each
category-year group. These were considered the top-cited articles in
the category-year group. We determined how many of these top-cited
articles were published in non-elite journals. 

Given the large number of subject categories under study (261), we
grouped subject categories into broad research areas. We used the
broad areas from Scholar Metrics for this, with one change --- we
separated {\it Engineering} and {\it Computer Science}. The citation
patterns in these two areas are significantly different and this
separation allowed us to explore the differences. We also added {\it
  All articles} as the union of all broad areas.

For each year, we sorted the subject categories in each broad area by
the number of top-cited articles published in non-elite journals. We
then picked the median subject category in each broad area as the
representative for the area --- roughly half the subject categories in
the area would have more top-cited articles published in non-elite
journals than the representative and roughly half would have fewer
such articles. Note that we recomputed the representative category for
each year to ensure that we picked the middle point in the list of
subject categories at all times.

We picked the median subject category as the representative instead of
computing an average across all categories to limit distortions due to
outliers. In addition to selecting a representative category, we also
computed the 25th and the 75th percentile categories for each broad
area in each year. These are the categories whose number of non-elite
top-cited articles was larger than or equal to that for 25\% and 75\%
of all the categories in the area, respectively. The results for these
categories help us get an idea of how the number of top-cited articles
in non-elite journals varied across the entire set of subject
categories in an area.

Finally, we computed the number of citations to all articles in a
category-year group as well as the number of citations to all articles
in the group that were published in non-elite journals.

\section{Results}

\begin{figure}
\centering
\includegraphics{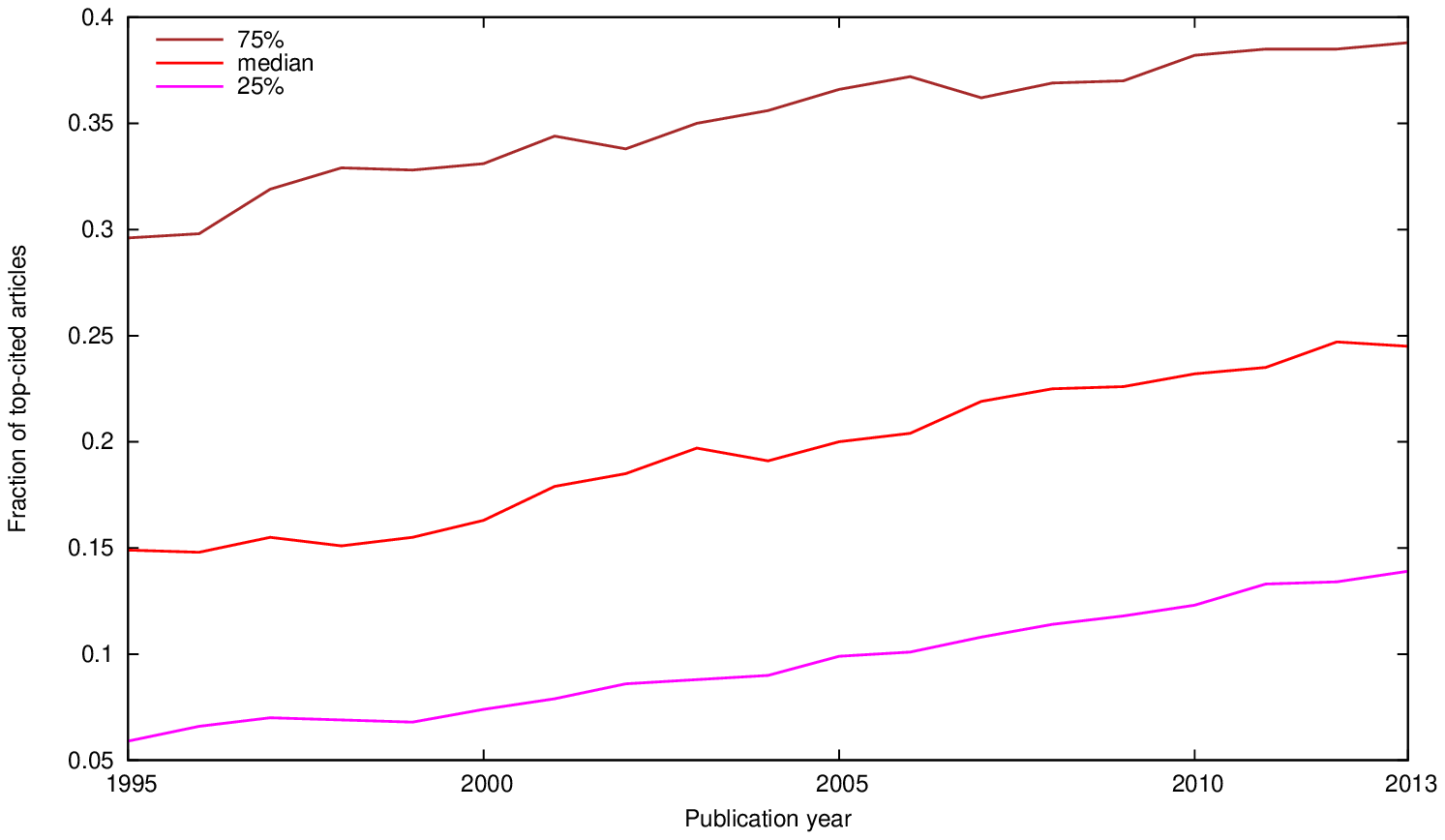}\\
{\small The line marked {\it median} presents the data for the representative
  subject category in each year.  Lines marked 25\% and 75\% represent
  results for the 25th and 75th percentile categories, respectively.}
\caption{Fraction of top-cited papers published in non-elite journals over 1995-2013.}
\label{fig:all-top-cited} 
\end{figure}

Figure~\ref{fig:all-top-cited} presents the trend in publication of
top-cited papers. It shows that the fraction of top-cited papers
published in non-elite journals has grown steadily over 1995-2013. The
graphs for the representative subject category as well as those for
the 25th and 75th percentile categories are similar. This indicates
that the trend of growth in fraction of top-cited papers published in
non-elite journals holds across a wide range of subject categories.

Figure~\ref{fig:broad-topcited-1} presents the trend in publication of
top-cited papers for individual broad areas. It shows that all areas
saw growth in the fraction of top-cited papers published in non-elite
journals; 6 out of 9 broad areas seeing a substantial increase.

Table~\ref{tab:broad-topcited} presents the number of 2013 top-cited
papers published in non-elite journals for the representative
subject category in each broad area. It also presents the change since
1995. The change over 1995-2013 is computed as a percentage:\\

$(num\_non\_elite\_in\_2013-num\_non\_elite\_in\_1995)/num\_non\_elite\_in\_1995 * 100$\\

It shows that the number of top-1000 papers published in non-elite
journals for the overall representative subject category went from 149
in 1995 to 245 in 2013, a growth of 64\%. Four out of nine broad areas
saw at least one-third of the top-cited articles published in
non-elite journals in 2013. All broad areas of research saw a growth
in the fraction of top-cited articles published in non-elite journals
over 1995-2013. For 6 out of 9 broad areas, the fraction of top-cited
papers published in non-elite journals for the representative subject
category grew by 45\% or more.

\begin{table}[htbp]
\centering
\begin{tabular}{|l|r|r|r|}
\hline
Broad area & Top-cited non-elite in 2013 & Change since 1995 \\  \hline
Physics \& Mathematics & 289 & 204\%\\
Health \& Medical Sciences & 192 & 98\%\\
Chemical \& Material Sciences & 108 & 80\%\\
Computer Science & 345 & 72\%\\
Engineering & 174 & 63\%\\
Business, Economics \& Management & 333 & 45\%\\
Social Sciences & 349 & 18\%\\
Life Sciences \& Earth Sciences & 177 & 18\%\\
Humanities, Literature \& Arts & 414 & 6\%\\
{\it All articles} & {\it 245} & {\it 64\%}\\
\hline
\end{tabular}\\[0.25cm]
{\small The numbers for each broad area are from the representative 
subject category for the area.
}
\caption{Change in the number of top-1000 papers published in non-elite journals over 1995-2013. }
\label{tab:broad-topcited}
\end{table}

\begin{figure}
\centering
\subfloat[{\small Physics \& Mathematics}]{\includegraphics{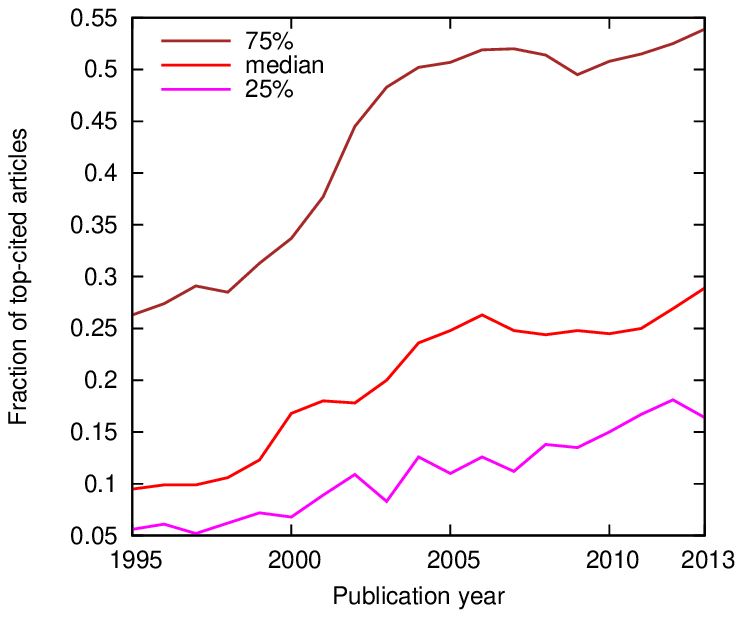}}\,
\subfloat[{\small Health \& Medical Sciences}]{\includegraphics{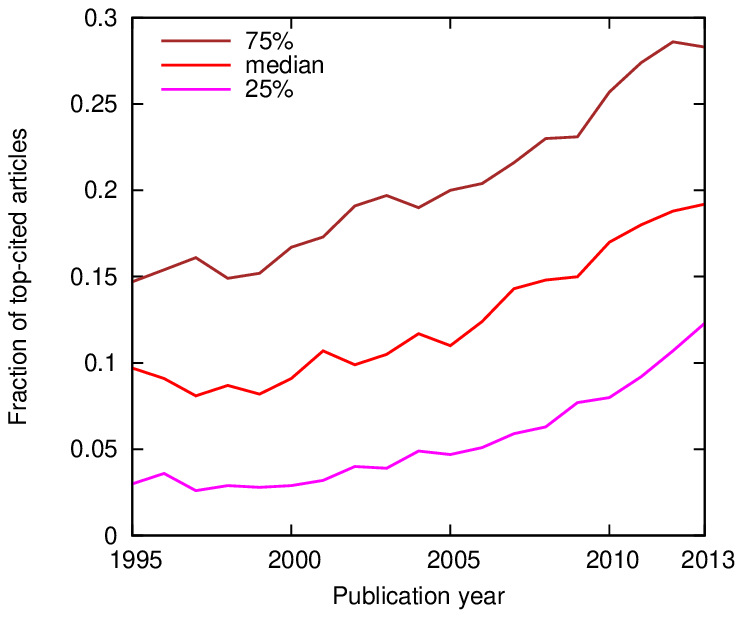}}\,
\subfloat[{\small Chemical \& Material Sciences }]{\includegraphics{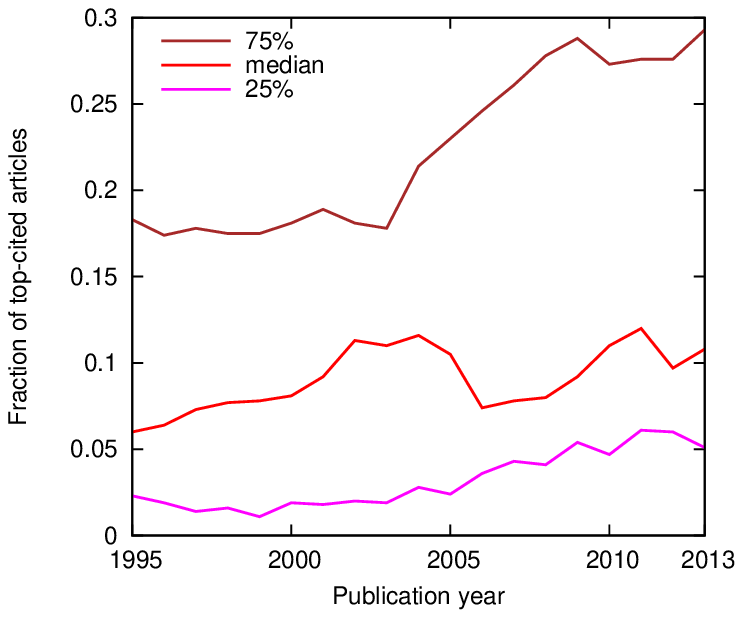}}\,
\subfloat[{\small Computer Science}]{\includegraphics{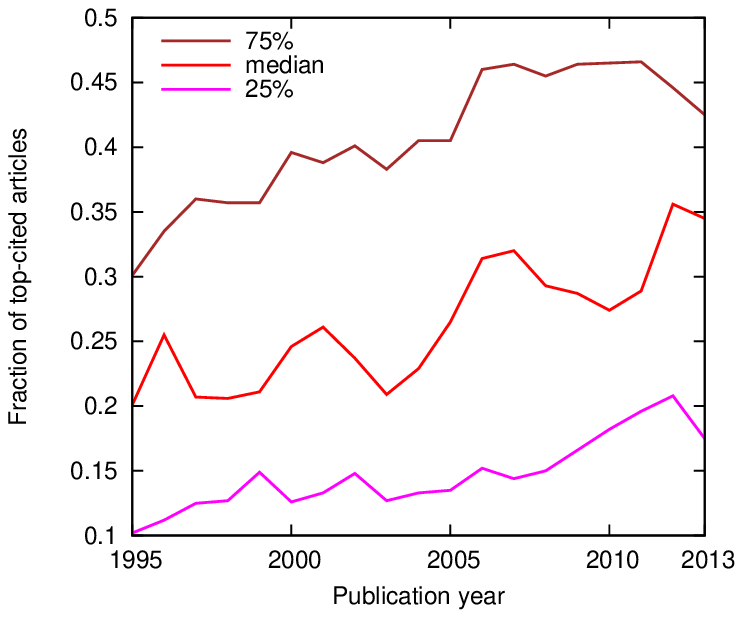}}\,
\subfloat[{\small Business, Economics \& Management}]{\includegraphics{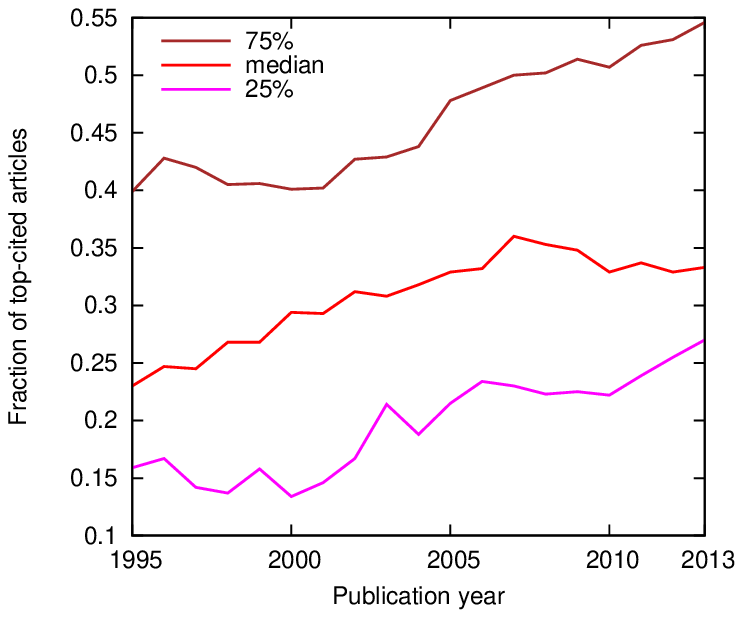}}\,
\subfloat[{\small Engineering} ]{\includegraphics{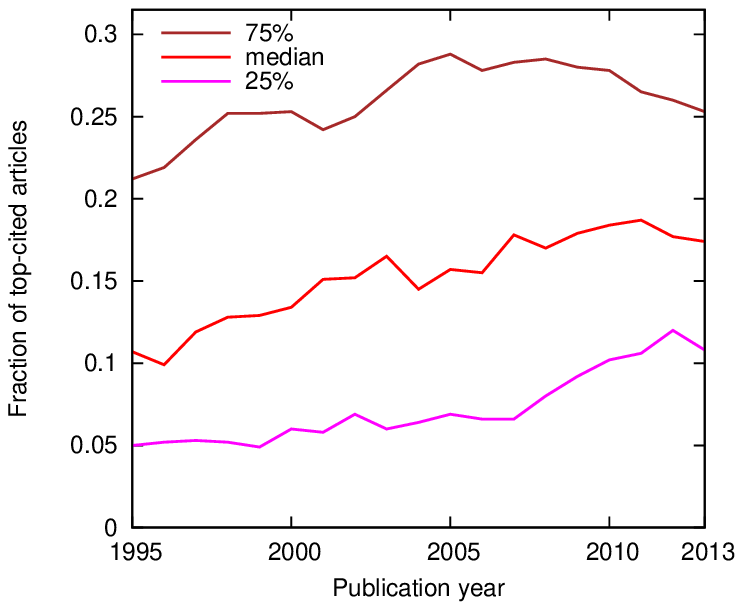}}\,
\caption{Per-area changes in the fraction of top-cited papers published in non-elite journals over 1995-2013. }
\label{fig:broad-topcited-1} 
\end{figure}

\begin{figure}
\ContinuedFloat
\centering
\subfloat[{\small Social Sciences}]{\includegraphics{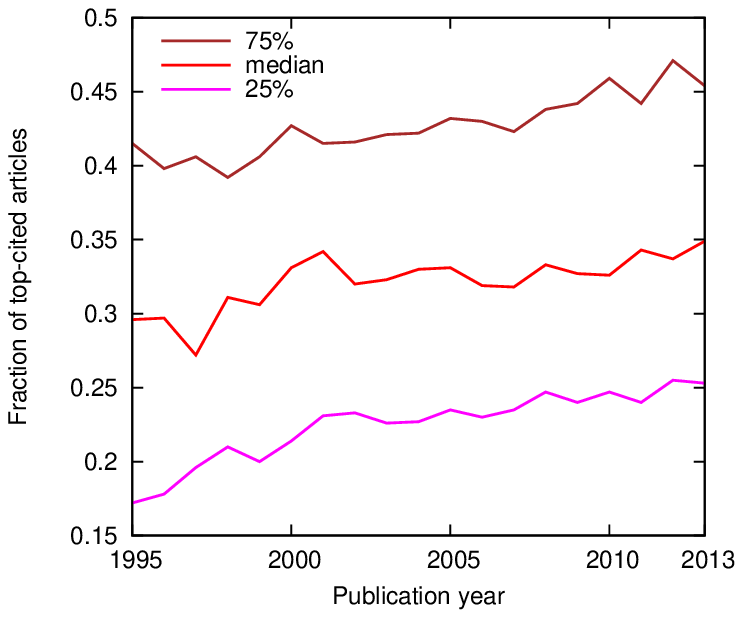}}\,
\subfloat[{\small Life Sciences \& Earth Sciences }]{\includegraphics{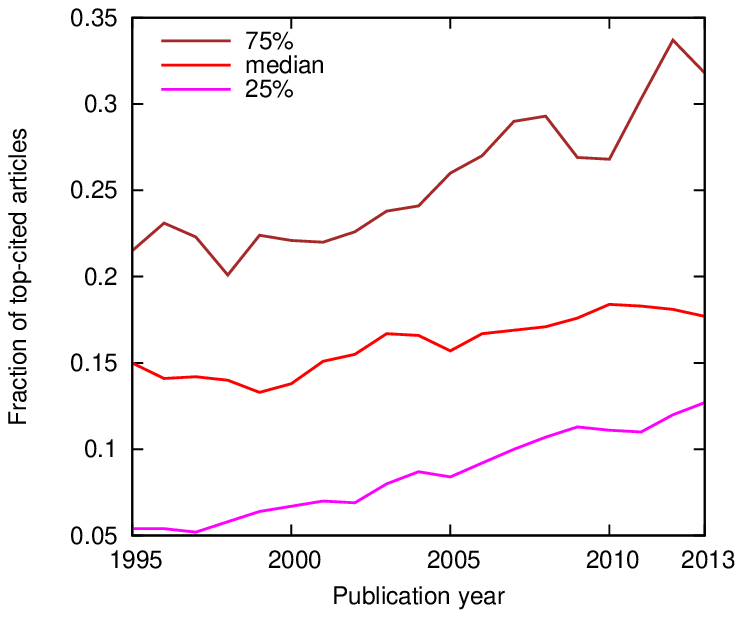}}\,
\subfloat[{\small Humanities, Literature \& Arts}]{\includegraphics{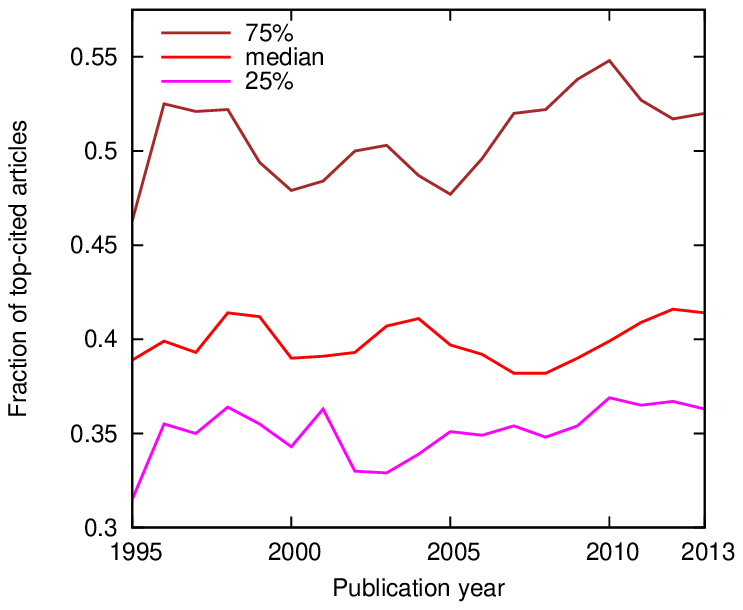}}\,
\caption{Per-area changes in the fraction of top-cited papers published in non-elite journals over 1995-2013.}
\end{figure}

\begin{figure}
\centering
\includegraphics{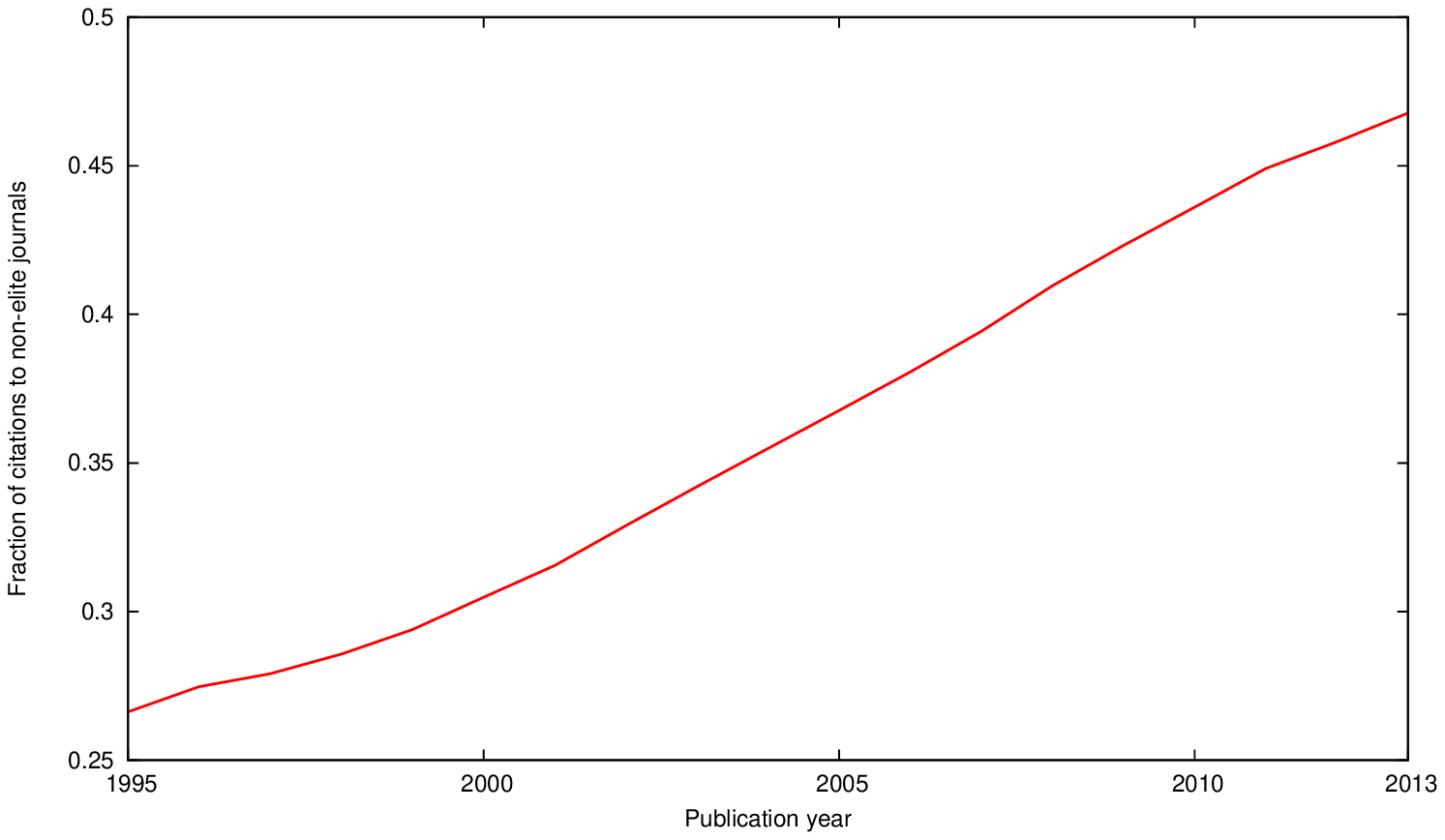}\\
\caption{Fraction of citations to non-elite journals over 1995-2013.}
\label{fig:all-total} 
\end{figure}

Figure~\ref{fig:all-total} presents the fraction of citations to
non-elite journals over 1995-2013. Note that this is computed by
dividing the sum of citations to all articles in non-elite journals by
the sum of citations to articles in all journals. The fraction of
non-elite citations went from 27\% in 1995 to 47\% in 2013. The graph
shows growth over the entire period, the growth rate being lower in
the first third of the period under study (1995-2000) and higher over
the rest of the period (2001-2013).

\begin{figure}
\centering
\includegraphics{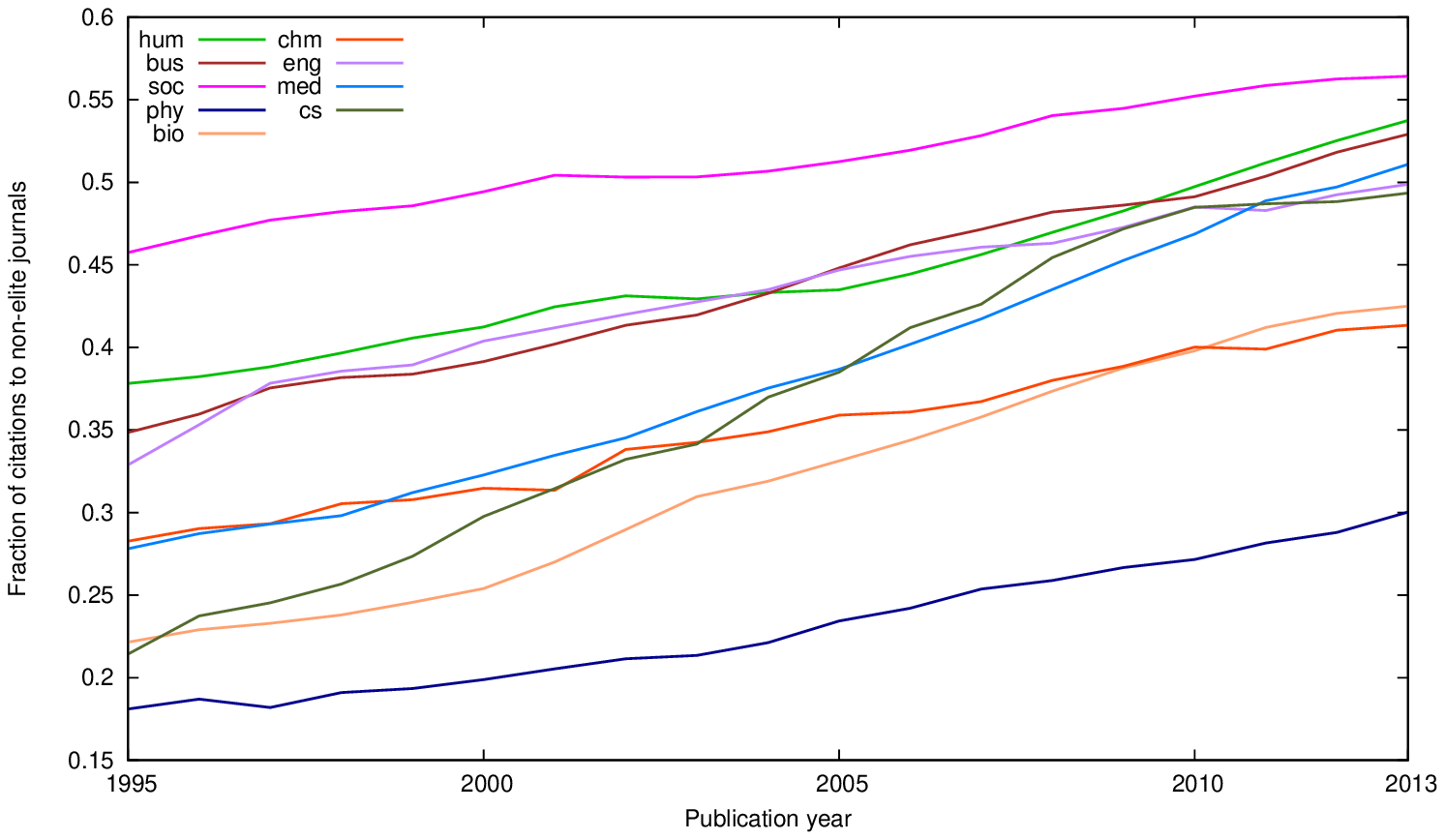} \\
{\small
{\bf bio:} Life Sciences \& Earth Sciences;
{\bf bus:} Business, Economics \& Management;
{\bf cs:} Computer Science;
{\bf chm:} Chemical \& Material Sciences;
{\bf eng:} Engineering;
{\bf hum:} Humanities, Literature \& Arts;
{\bf med:} Health \& Medical Sciences;
{\bf phy:} Physics \& Mathematics;
{\bf soc:} Social Sciences}\\
\caption{Fraction of citations to non-elite journals for broad areas of research. }
\label{fig:broad-total} 
\end{figure}

Figure~\ref{fig:broad-total} presents the evolution of the fraction of
non-elite citations for all broad areas. It shows that all broad areas
saw a significant increase in the fraction of non-elite citations over
1995-2013. 

\pagebreak

Table~\ref{tab:broad-frac-and-growth} presents the per-area
growth in numerical form, for ease of comparison. It shows that six
out of nine broad areas had at least 50\% of total citations going to
articles published in non-elite journals in 2013. Furthermore, 8 out
of 9 broad areas saw an increase of 40\% or more in the fraction of
citations to non-elite journals over 1995-2013.

\begin{table}
\centering
\begin{tabular}{|l|r|r|r|}
\hline
Broad area & Non-elite citations in 2013 & Change since 1995 \\  \hline
Computer Science & 50\% & 133\%\\
Life Sciences \& Earth Sciences & 43\% & 95\%\\
Health \& Medical Sciences & 51\% & 83\%\\
Physics \& Mathematics & 30\% & 67\%\\
Engineering & 50\% & 52\%\\
Business, Economics \& Management & 53\% & 51\%\\
Chemical \& Material Sciences & 41\% & 46\%\\
Humanities, Literature \& Arts & 53\% & 42\%\\
Social Sciences & 56\% & 22\%\\
{\it All articles} & {\it 47\%} & {\it 74\%}\\
\hline
\end{tabular}
\caption{Change in the fraction of citations to non-elite journals over 1995-2013.}
\label{tab:broad-frac-and-growth}
\end{table}

\section{Related Work}

The idea that a small core set of journals covers most of the key
papers in a discipline has long been prevalent in the study of
scholarly communication. For example,
see~\cite{bradford1934sources,garfield1972citation,garfield1996significant}
or do a query for ``core journals'' in article titles on Google
Scholar~\cite{scholar-core-journals2014}.

Larivi{\`e}re et~al~\cite{lariviere2009decline} examined the
concentration of citations at the article level over 1900-2007 and
concluded that while distributions of citations remained highly
skewed, the fraction of citations to highly cited articles has been
decreasing. Looking at broad research areas, they found that for {\it
  Natural Sciences \& Engineering} and {\it Medicine}, the decrease in
concentration started around 1990.

Lozano et~al~\cite{lozano2012weakening} studied the relationship
between impact factor and article citations. They examined three broad
categories, {\it Natural Sciences \& Medicine}, {\it Physics} and {\it
  Social Sciences}, over 1902-2009 and found that the correlation
between impact factor of the journal and the citations its articles
receive has been weakening since around 1990. In another part of their
study, they computed the fraction of 10\% most-cited articles that
appear in the 10\% highest impact-factor journals. For this, they
focused on {\it Natural Sciences \& Medicine}. They found that a
decreasing fraction of most-cited articles is being published in the
most-cited journals.

In a follow-up study~\cite{lariviere2014elite}, they took a closer
look at 13 journals, seven traditionally top-ranked journals and six
upcoming journals. They found that since around 1990, the fraction of
most-cited papers published in the traditionally top-ranked journals
has been dropping and the fraction of such papers published in the
upcoming journals has been increasing.

The goal of our study is similar to those of Larivi{\`e}re
et~al~\cite{lariviere2014elite, lozano2012weakening}. We have examined
the impact of non-elite journals for a large number of specific
research fields. Structuring such a study to consider individual
research fields separately is necessary since: (1) usually only the
journals in a specific field are the possible venues for the
highly-cited articles in the field, (2) there are large differences in
the citation frequency in different fields and grouping a large number
of research fields usually results in the fields with higher citation
frequency dominating the results. That said, our results are
complementary to theirs and point in the same direction.

\section{Conclusions}

There are two main conclusions from our study. First, the fraction of
highly-cited articles published in non-elite journals increased
steadily over 1995-2013. While the elite journals still publish a
substantial fraction of high-impact articles, many more authors of
well-regarded papers in a diverse array of research fields are
choosing other venues.

Our analysis indicates that the number of top-1000 papers published in
non-elite journals for the representative subject category went from
149 in 1995 to 245 in 2013, a growth of 64\%. Looking at broad
research areas, 4 out of 9 broad areas saw at least one-third of the
top-cited articles published in non-elite journals in 2013. All broad
areas of research saw a growth in the fraction of top-cited articles
published in non-elite journals over 1995-2013. For 6 out of 9 broad
areas, the fraction of top-cited papers published in non-elite
journals for the representative subject category grew by 45\% or more.

Second, the fraction of citations to articles published in non-elite
journals has grown substantially over most research areas. The
percentage of citations to articles in non-elite journals went from
27\% of all citations in 1995 to 47\% in 2013. Six out of nine
categories had at least 50\% of total citations going to articles
published in non-elite journals in 2013.

Now that finding and reading relevant articles in non-elite journals
is about as easy as finding and reading articles in elite journals,
researchers are increasingly building on and citing work published
everywhere.

\bibliographystyle{plain}
\bibliography{paper}

\end{document}